\documentclass[sigconf,nonacm]{acmart}

\renewcommand\footnotetextcopyrightpermission[1]{}
\usepackage{xcolor}

\begin{document}

\title{SyncSpace: Layout-Conditioned 3D Gaussian Splatting for Space Reskinning in Mixed Reality
}

\author{Qinchuan Zhang}
\email{qinchuan@usc.edu}
\affiliation{%
  \institution{University of Southern California}
  \city{Los Angeles}
  \state{California}
  \country{USA}}

\author{Weibo Xu}
\email{weiboxu@usc.edu}
\affiliation{%
  \institution{University of Southern California}
  \city{Los Angeles}
  \state{California}
  \country{USA}}

\author{Yunge Wen}
\email{yungew@mit.edu}
\affiliation{%
  \institution{Massachusetts Institute of Technology}
  \department{MIT Media Lab}
  \city{Cambridge}
  \state{Massachusetts}
  \country{USA}}

\renewcommand{\shortauthors}{Zhang et al.}

\begin{abstract}
We present \emph{SyncSpace}, a system that achieves both spatial alignment and visual consistency between a generated 3DGS world and physical space. We first scan the space via depth sensing to extract 3D bounding boxes, which we render into a layout-only panorama and feed as a geometric prior to a generative world model, producing a Gaussian splat scene in which objects are re-semantized to fit a target style without per-object control. We then align the generated scene to physical space with a coarse-to-fine registration algorithm, refined manually via pinch gestures when automatic registration does not converge. We demonstrate a hand-tracked engulfment interaction in which the virtual world rises to replace the physical space, and show a single space reskinned into multiple stylistically distinct worlds with its layout preserved.
\end{abstract}

\begin{teaserfigure}
  \includegraphics[width=\textwidth]{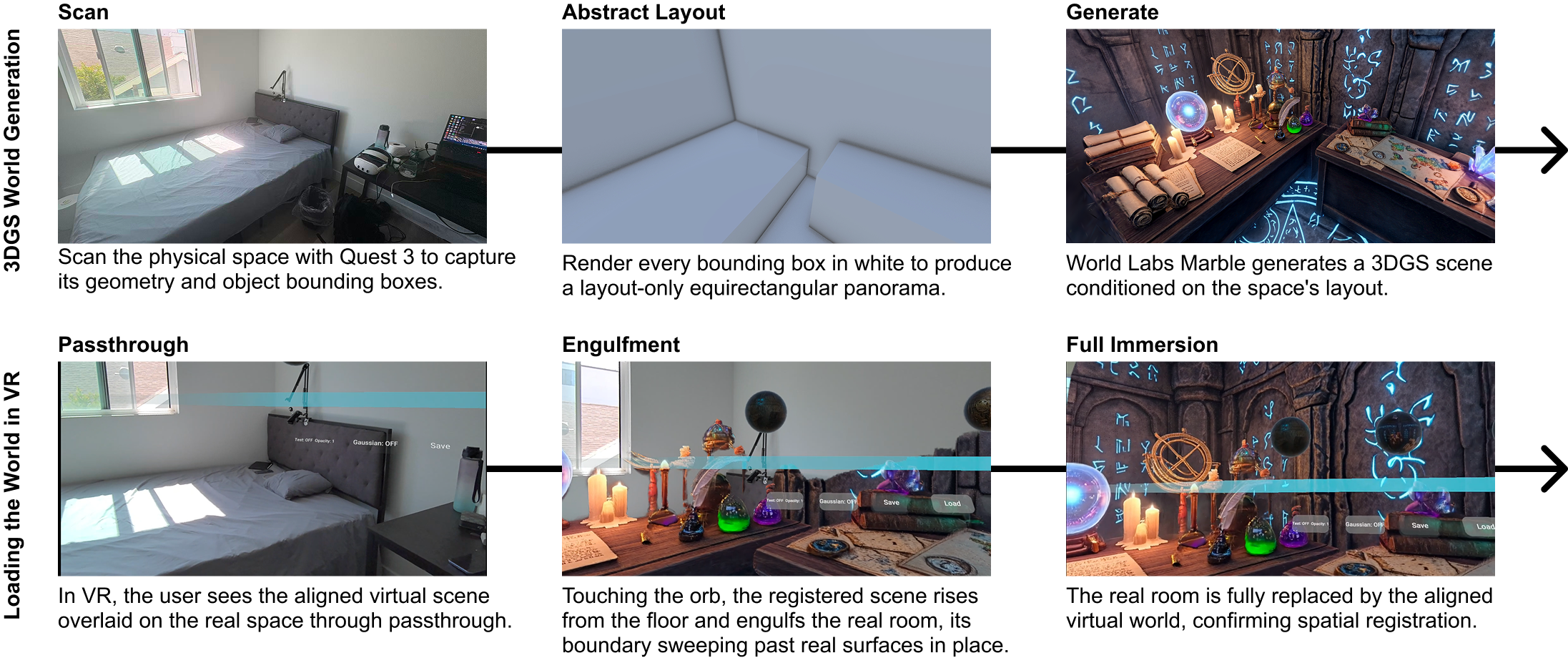}
  \caption{\textbf{SyncSpace} reskins a real room into a stylistically unified,
  spatially congruent MR world in minutes. \emph{Top:} the generation
  pipeline---scan the room, abstract it to a layout-only white panorama, and
  generate a layout-conditioned Gaussian-splat world. \emph{Bottom:} the in-VR
  experience---touching a virtual orb makes the virtual world rise and engulf the
  real room until it fully replaces it, preserving layout while re-imagining
  appearance.}
  \Description{A six-panel figure in two rows. The top row shows the generation
  pipeline: a scanned real bedroom, the same room abstracted to white
  layout-only geometry, and a stylized Gaussian-splat wizard-study world. The
  bottom row shows the in-VR first-person experience: the real room in
  passthrough, the virtual world rising from the floor to engulf it, and the
  fully replaced spatially congruent virtual world.}
  \label{fig:teaser}
\end{teaserfigure}

\maketitle

\section{Introduction}

Mixed reality headsets can now render immersive, room-scale virtual worlds in real time, enabling gaming, creative exploration, and themed entertainment at home. Depth-sensor-based 3D scanning can reliably extract a space's walkable range and prevent collisions via geometric shape matching, but generating a virtual world that is both stylistically unified and registered to this scan remains an open problem.

Substitutional-reality and proxy-based systems \cite{Simeone2015,Hettiarachchi2016} match individual real objects to shape-similar virtual props, while layout-driven approaches \cite{Sra2016,Cheng2019,Yang2019} instead retrieve entire pre-authored room templates to fit real-world geometry, constructing walkable VR at the scale of the whole space. Both are constrained to whatever their asset library contains, producing stylistically inconsistent results. The emergence of 3D generative models has enabled recent style-transformation systems \cite{Yang2024DreamSpace,Wang2026Roomify} to replace retrieval with generation, such as diffusion-based texture propagation onto scanned meshes or per-object 2D-to-3D conversion, yet global style coherence is still imposed post hoc rather than emerging intrinsically from the generative process. 3D Gaussian Splatting \cite{Kerbl2023} has further enabled real-time, photorealistic rendering, allowing generative world models \cite{WorldLabs2024} to synthesize an entire room in a single generative pass, bypassing per-object assembly altogether; however, the resulting content is unregistered to the user's physical space, precluding safe walking. No existing approach simultaneously achieves whole-room stylistic coherence, spatial registration to the physical environment, and safe embodied navigation.

We present \emph{SyncSpace}, a system that achieves both spatial alignment and visual consistency between a generated 3DGS world and physical space. We first scan the space via depth sensing to extract 3D bounding boxes, which we render into a layout-only panorama and feed as a geometric prior to a generative world model, producing a Gaussian splat scene in which objects are re-semantized to fit a target style without per-object control. We then align the generated scene to physical space with a coarse-to-fine registration algorithm, enabling safe embodied navigation through a stylistically transformed environment.

\section{SyncSpace}

SyncSpace decouples a physical layer (bounding boxes providing per-object colliders and occlusion) from a visual layer (the GS scene, supplying appearance only), so that reskinning the space's appearance never compromises safe navigation.

\subsection{World Generation}

\textbf{Physical Space Scanning \& Abstraction.} SyncSpace begins by scanning the space with a Meta Quest~3/3S, which produces axis-aligned bounding boxes for all detected objects and surfaces via depth-sensor-based 3D scanning. We render every box with a uniform white material, stripping object identity while retaining spatial extent. This gives the world model a geometry-only signal, so whole-room style is applied unconstrained by existing object semantics. The output is a white equirectangular panorama at $4096\times2048$ encoding spatial layout alone.

\textbf{Layout-Conditioned 3DGS Generation.}
The white-rendered panorama is paired with a whole-room style prompt (e.g., \emph{Enchanted Wizard's Study}) and sent to Marble~1.1 (World Labs), which generates a Gaussian splat scene \cite{Kerbl2023} conditioned on the space's geometric layout. Unlike per-object prompting, the prompt describes a single coherent aesthetic for the entire space, so the model reinterprets every object's appearance within that style without any object-level control. As a result, objects are \emph{emergently re-semantized}: their visual identity shifts to match the target world (Fig.~\ref{fig:styles}) while their spatial positions remain fixed.

\subsection{Spatial Alignment}

\textbf{Scale Alignment.} The generated GS scene and the room scan exist at different metric scales, because the world model has no absolute scale reference. We first rescale the GS scene so its axis-aligned bounding box matches that of the room scan, yielding a uniform scale factor $s$ applied to all Gaussian centers.

\textbf{Coarse-to-Fine Registration.} We register the two point clouds with Open3D in two stages. We first extract FPFH descriptors for each point and estimate a coarse rigid transformation $(R, t)$ via RANSAC, minimizing the correspondence error over inlier matches $\{(p_i, q_i)\}$:
\begin{equation}
(R^*, t^*) = \arg\min_{R,t} \sum_{i} \| (R p_i + t) - q_i \|^2
\end{equation}
where $R \in SO(3)$ and $t \in \mathbb{R}^3$. This coarse pose is then refined with point-to-plane ICP, which minimizes the distance from each source point to the local tangent plane of its nearest target point $q_i$ with normal $n_i$:
\begin{equation}
(R^*, t^*) = \arg\min_{R,t} \sum_{i} \left[ (R p_i + t - q_i) \cdot n_i \right]^2
\end{equation}
The transformation is updated iteratively until convergence, producing a registered GS scene in the physical coordinate frame.

\textbf{Manual Refinement for Imperfect Registration.} Automatic registration may not fully converge due to noisy generated geometry: single-panorama reconstruction is ill-posed, so occluded regions are hallucinated rather than measured \cite{Eigen2014}, and 3D Gaussian Splatting optimizes photometric rather than geometric fidelity, producing floaters and off-surface spread even in well-observed regions \cite{Kerbl2023,Huang2024_2DGS,Guedon2024SuGaR}. We therefore design an interface that allows users to refine the pose with pinch gestures, rendering the GS scene at 20\% opacity over passthrough for side-by-side comparison, until satisfactory alignment is reached.

To further demonstrate spatial congruence, we implement a hand-tracked engulfment interaction: the user touches a virtual orb, and the reskinned scene rises from the floor to engulf the real room, with the virtual floor meeting the real floor and virtual objects occupying the volume of their physical sources. Because the scene is registered to the room, any misalignment is immediately visible during this transition.

\section{Experiments}
\begin{figure}[t]
  \centering
  \includegraphics[width=\columnwidth]{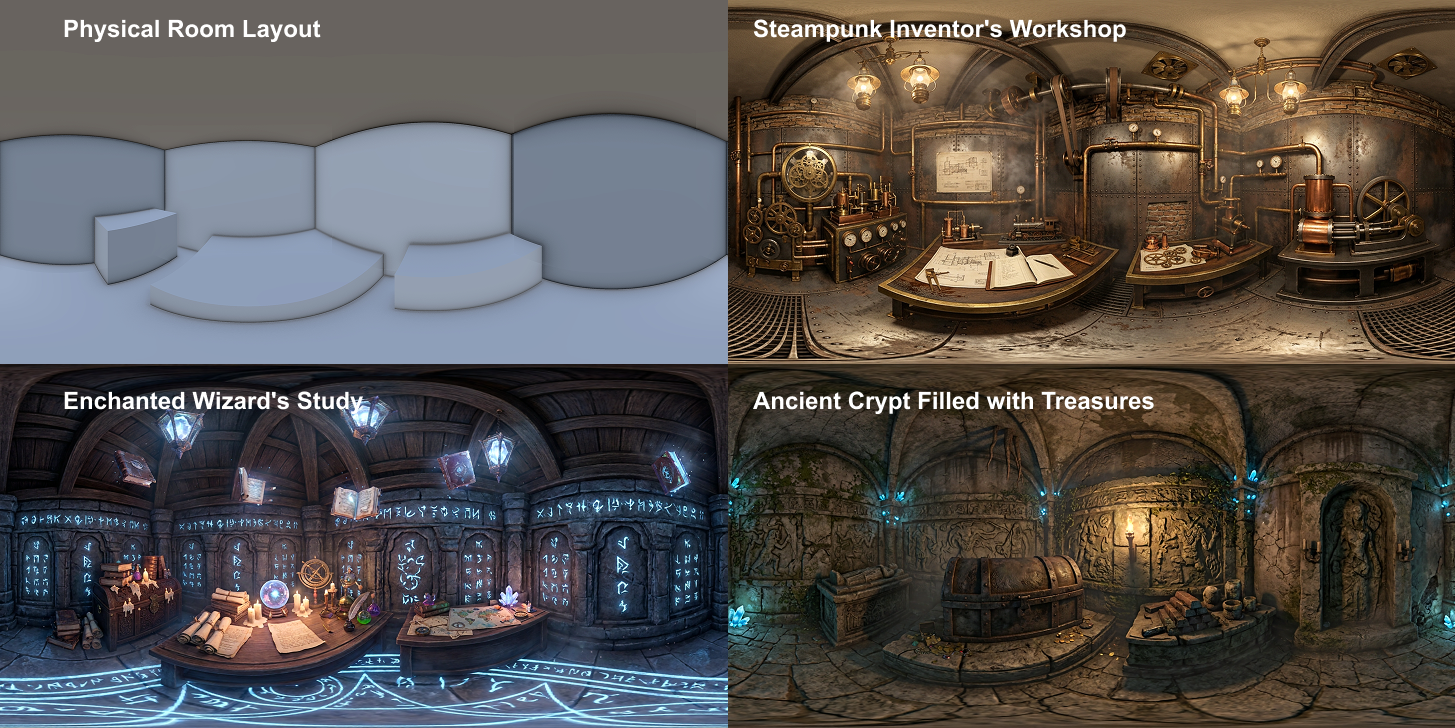}
  \caption{The same room reskinned under three style prompts, each preserving the layout while re-semantizing every surface (e.g., the same desk becomes a workbench, an alchemy table, or a treasure chest).}
  \Description{A two-by-two grid of equirectangular panoramas of the same room.
  Top-left is a white layout-only abstraction; the other three show the room
  restyled as a steampunk inventor's workshop, an enchanted wizard's study, and
  an ancient treasure crypt, all preserving the same spatial layout.}
  \label{fig:styles}
\end{figure}

We recruited three users, each of whom generated a distinct style for the same room (Fig.~\ref{fig:styles}: \emph{Enchanted Wizard's Study}, \emph{Ancient Crypt Filled with Treasures}, and \emph{Steampunk Inventor's Workshop}), completing the full pipeline in an average of 10 minutes. All three produced stylistically unified worlds that preserved the room's layout while re-semantizing its objects in markedly different ways, isolating the style prompt's effect on a fixed layout. A formal user study and multi-room generalization are left to future work.

\section{Discussion \& Future Work}

SyncSpace shows that layout-conditioned generation enables style-consistent, spatially safe MR world creation without modeling skill. Limitations remain: re-semantization is controllable only at the whole-space level, not per object; the GS scene is static and non-editable; automatic registration may need manual refinement when geometry is noisy. Future work includes per-object controllable generation, dynamic objects, cross-platform support, outdoor scenes, a formal user study, and eliminating manual alignment as world models improve.

\bibliographystyle{ACM-Reference-Format}
\bibliography{references}

\end{document}